\documentclass[journal]{IEEEtran}

\usepackage{epsfig}
\usepackage{graphicx}
\usepackage{amssymb}
\usepackage{mathtools}
\usepackage{caption}
\usepackage{subcaption}
\usepackage{hyperref}
\usepackage{multirow}
\usepackage{xcolor}

\newcommand{\beq}{\begin{equation}}
\newcommand{\enq}{\end{equation}}

\title{Application Space and the Rate-Distortion- Complexity Analysis of Neural Video CODECs}

\author{
	Ricardo L. de Queiroz, {\em Fellow, IEEE},
	Diogo C. Garcia, {\em Senior Member, IEEE},
	Yi-Hsin Chen, {\em Student Member, IEEE},
	Ruhan Concei\c{c}\~{a}o, {\em Student Member, IEEE},
	Wen-Hsiao Peng, {\em Fellow, IEEE},
        Luciano V. Agostini {\em Senior Member, IEEE}.
\thanks{R. L. de Queiroz is with the Computer Science Department at Universidade de Bras\'{i}lia, Brazil. 
D.C. Garcia is with the Faculty of Engineering Sciences and Technology at Universidade de Bras\'{i}lia, Brazil, 
Y.-H. Chen is with the Computer Science Department at National Yang Ming Chiao Tung University, Taiwan. 
R. Concei\c{c}\~{a}o is with the Center of Technological Development at Universidade Federal de Pelotas, Brazil. 
W.-H. Peng is with the Computer Science Department at National Yang Ming Chiao Tung University, Taiwan. 
L. V. Agostini is with the Center of Technological Development at Universidade Federal de Pelotas, Brazil. 
}
}

\begin{document}

\maketitle

\begin{abstract}
We study the decision-making process for choosing video compression systems through a rate-distortion- complexity (RDC) analysis. 
We discuss the 2D Bjontegaard delta (BD) metric and formulate generalizations in an attempt to extend its notions to the 3D RDC volume. 
We follow that discussion with another one on the computation of metrics in the RDC volume, and on how to define and measure the cost of a coder-decoder (codec) pair, where the codec is characterized by a cloud of points in the RDC space. 
We use a Lagrangian cost $D+\lambda R + \gamma C$, such that choosing the best video codec among a number of candidates for an application demands selecting appropriate $(\lambda, \gamma)$ values. 
Thus, we argue that an application may be associated with a $(\lambda, \gamma)$ point in the application space.  
An example streaming application was given as a case study to set a particular point in the $(\lambda, \gamma)$ plane. 
The result is that we can compare Lagrangian costs in an RDC volume for different codecs for a given application.  
Furthermore, we can span the plane and compare codecs for the entire application space filled with different $(\lambda, \gamma)$ choices.
We then compared several state-of-the-art neural video codecs using the proposed metrics. 
Results are informative and surprising. 
We found that, within our RDC computation constraints, only four neural video codecs came out as the best suited for any application, depending on where its desirable $(\lambda, \gamma)$ lies.
\end{abstract}
%
%

\section{Introduction} 

The rate-distortion (RD) analysis of video coder-decoder (codec) pairs is commonplace. However, video codec complexity is usually measured in execution time and decisions are commonly subjective, as they usually take into consideration other factors that affect the cost of a codec. 
These factors 
can be as \textit{direct} as its rate, distortion, manufacturing, operation and maintenance costs, 
or \textit{indirect} such as the cost to rewrite the specs, replace the code base, replace equipments, etc. They can even be \textit{subjective} such as the need to graduate, publish, or to maintain a brand afloat. 
In standards or industry, an idea that improves a codec can be rejected for being ``too costly for so little gain'', but it is not common to quantify the complexity axis in a way that promotes a thorough rate-distortion-complexity (RDC) analysis, which could lead to prompt adoption decisions. 

There are limited ways to measure the rate in coding applications. Typically, it is quantified as the number of bits or bytes required to transmit a video per second, which can be expressed in units such as kilobytes per second (kB/s), megabytes per second (MB/s), kilobits per second (kb/s), or other similar metrics.

Distortion and quality are represented in current works using various methods. Sum of Absolute Differences (SAD) calculates the sum of absolute differences between corresponding pixel values in two images. It is primarily used in motion estimation \cite{OverviewVVC, HEVC_Comparison} for video encoding rather than for evaluating image quality. The most common method to evaluate the distortion caused by a codec is through Mean Squared Error (MSE), which measures the average squared intensity differences between the reference and test frames. 

Another widely used metric to measure quality is Video Multi-Method Assessment Fusion (VMAF) \cite{VMAF}. VMAF combines multiple quality metrics, including human visual perception models, and uses machine learning to predict the viewer's satisfaction. It generates a single score typically between 0 and 100, where higher scores indicate better visual quality. VMAF has become a standard for evaluating video compression and streaming performance.

Complexity and the overall cost associated with codecs will be discussed at a later section. 

In this paper, we want to propose methods to perform RDC analysis, so that different codecs can be compared on a wide range of applications. 

\section{Related Works}

Herglotz \textit{et. al.} provided a comprehensive overview of the BD calculus and its existing implementations~\cite{herglotz2024bjontegaard}, while presenting an analysis of distinct BD interpolation algorithms.
In addition, the authors also propose a metric called Relative Curve Difference (RCD) that can represent the actual savings at a fixed quality point. 

Katsenou \textit{et. al.}~\cite{katsenou2022energy} explore and compare energy consumption in both encoding and decoding processes across optimized, state-of-the-art traditional video codecs. 
Additionally, the authors propose the metric Energy-to-Bitrate (EBR), which quantifies the energy consumed per unit of bitrate. The perceptual quality of the video is evaluated using both PSNR and VMAF metrics. 

In another work \cite{kranzler2024comprehensive}, it was carried a comprehensive analysis of energy and compression efficiency across different video coding formats, evaluating both software and hardware decoding implementations. It was demonstrated that, for certain codecs, the hardware decoder can significantly reduce the energy demand, as compared to software decoding.

An objective evaluation method named Generalized Bjøntegaard Delta PSNR (GBD-PSNR) was also proposed~\cite{li2010rate}, which incorporates coding complexity as an additional dimension, resulting in a RDC evaluation surface.
Such a metric, however, has some limitations. For example, when comparing two codecs with very different complexities, often their RDC surfaces have minimal or no overlapping regions, thus limiting the metric effectiveness. 
Furthermore, the coding complexity is measured in terms of processing time, which is implementation- and platform-specific, as previously discussed.

The above works exclusively focus on traditional codecs, whereas this paper emphasizes neural video codecs. Furthermore, none of the previous works address practical considerations towards a comprehensive RDC analysis, as the present paper does.


\section{Complexity and the overall cost of a codec}

Complexity can be defined and characterized in many different ways. In traditional codecs, like H.264/AVC \cite{h264}, H.265/HEVC \cite{h265}, H.266/VVC \cite{h266}, AV1 \cite{av1}, etc., many works employ encoding and decoding runtimes as complexity metrics \cite{time_example1,time_example2, time_example3, time_example4,time_example5, time_example6, time_example7, time_example8}. However, using runtime to measure complexity has some limitations, since it highly depends on hardware and/or software optimizations, and environmental issues, thus making it a non-standardized approach for fair comparisons across different platforms.

For neural video codecs, an important metric is the number of multiply–accumulate  (MAC) operations, usually expressed in thousands of MAC per pixel (kMAC/pixel). 
This is an intrinsic complexity that is usually platform-independent. Other important metrics for neural video codecs are the memory bandwidth, decoded picture buffer size, or power consumption. 
One could argue that complexity is a multidimensional problem that would be better described as a vector of ``complexities''. This, however, would further complicate the analysis. 

What is the overall cost of a codec that we are trying to minimize? If we let the complexity measurements, along with rate and distortion, be grouped into a vector of cost metrics ${\bf s} = [s_0, s_1, \ldots, s_{K}]^T$, we are looking for a function that would map $\bf s$ to a cost $J$ as $J = f({\bf s})$. 
What is usually done (which includes standard bodies) is to apply a linear combination of rate ($R$) and distortion ($D$) and add a subjective analysis of complexity, guided by computation time. There is no repeatable evaluation process. 

Here, we are considering a linear approach to $f$, i.e. $\bf s$ is weighted by an importance vector ${\bf u} = [\lambda_0, \lambda_1, \ldots, \lambda_K]^T$ to yield the final cost as:

\begin{equation}
	J = {\bf u}^T {\bf s} .
\label{eq:linearcost}
\end{equation}

The implications are that minimizing $J$ for each possible $\bf u$ imply in spanning the lower convex hull of the space of all $\bf s$.
 
In this paper, we only consider $R$, $D$, and one form of complexity ($C$), such that ${\bf s}=[R,D,C]^T$.
Furthermore, for simplicity, we measure the complexity as the number of MAC. 
If the linear cost of RDC and the counting of kMAC/pixel is not deemed suitable for your application, our analysis can be easily adapted as long as the cost follows (\ref{eq:linearcost}).


\section{Some thoughts on BD metrics} 

\subsection{Preliminaries}
\label{sec:preliminaries}

A codec is usually judged based on its RD analysis, more specifically using the Bjontegaard delta (BD) metric \cite{bjontegaard2001}.
This metric is commonly calculated either as rate savings (BD-rate), expressed in percentage terms, or as an improvement in PSNR (BD-PSNR), which is measured in decibels (dB).

PSNR is a popular quality metric indirectly associated with distortion. It derives from mean squared error (MSE) as $PSNR = 10 \log_{10} \frac{255^2}{MSE}$ for 8-bit image data. 
We use the actual MSE distortion instead of PSNR in order to equally treat all three RDC dimensions, throughout this work. That is, we aim for minimizing the RDC cost along the three (or even more) dimensions. 

We also assume non-negative RDC values. 
However, if we compute the MSE in dB as $MSE^{dB}=10 \log_{10} MSE$, we have 

\begin{equation}
	PSNR_1 - PSNR_2 = MSE^{dB}_2 - MSE^{dB}_1 .
\end{equation}

\noindent 
In other words, a BD-PSNR measure is the negative of a BD-$MSE^{dB}$ measure. 
The latter is desirable because it allows us to minimize all RDC dimensions.  
Note that $MSE < 1$ implies a negative $MSE^{dB}$ and one needs to account for that. 
If image data is normalized to take on a value between 0 and 1, then $PSNR = -MSE^{dB}$, if not, there is a linear relation with an offset. 
For computing BD measurements, it is better to use $MSE^{dB}$ because of its direct relation to BD-PSNR. 
For optimization, however, it is better to use, instead, regular MSE, because the logarithm of only one axis can transform convex regions into non-convex.
What is important at this point is to consider that all $R$, $D$ and $C$ are non-negative, and that we want to minimize them all.

\subsection{BD computation}

\begin{figure}[tbp]
    \centering
    \begin{subfigure}{0.45\linewidth}
        \centering
        \includegraphics[trim={0 338px 389px 0},clip,width=\linewidth]{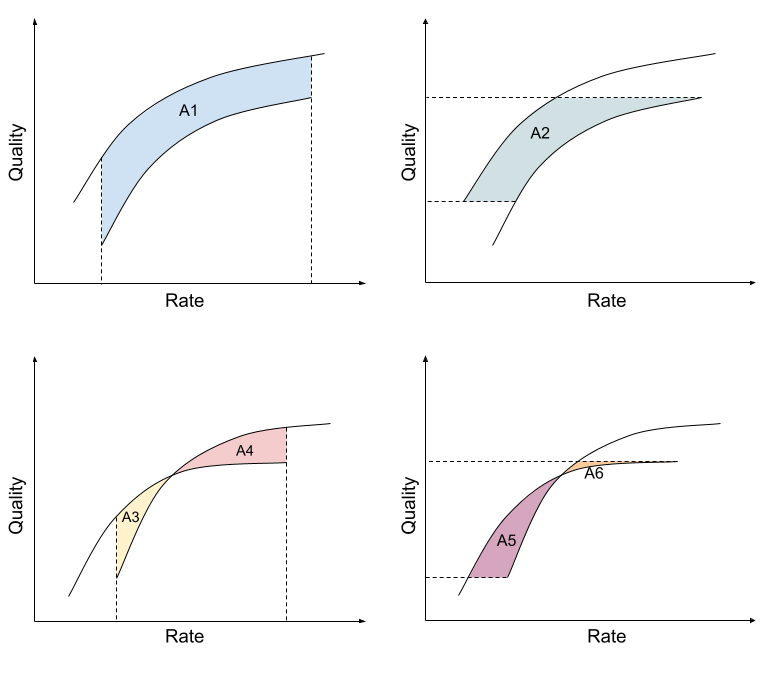}
        \caption{\ }
        \label{fig:bd1a}
    \end{subfigure}
    \begin{subfigure}{0.45\linewidth}
        \centering
        \includegraphics[trim={389px 338px 0 0},clip,width=\linewidth]{fig/queir_1.png}
        \caption{\ }
        \label{fig:bd1b}
    \end{subfigure}
    \begin{subfigure}{0.45\linewidth}
        \centering
        \includegraphics[trim={0 0 389px 338px},clip,width=\linewidth]{fig/queir_1.png}
        \caption{\ }
        \label{fig:bd1c}
    \end{subfigure}
    \begin{subfigure}{0.45\linewidth}
        \centering
        \includegraphics[trim={389px 0 0 338px},clip,width=\linewidth]{fig/queir_1.png}
        \caption{\ }
        \label{fig:bd1d}
    \end{subfigure}
    \caption{\ Illustration of BD-PSNR (a) as area $A_1$ and BD-rate (b) as area $A_2$. RD curves in (c) and (d) demonstrate crossing scenarios where BD-PSNR and BD-rate yield contradictory results due to differences in areas $A_3$, $A_4$, $A_5$ and $A_6$.}
    \label{fig:bd1}
\end{figure}

\begin{figure}[tbp]
    \centering
    \begin{subfigure}{0.45\linewidth}
        \centering
        \includegraphics[trim={0 338px 389px 0},clip,width=\linewidth]{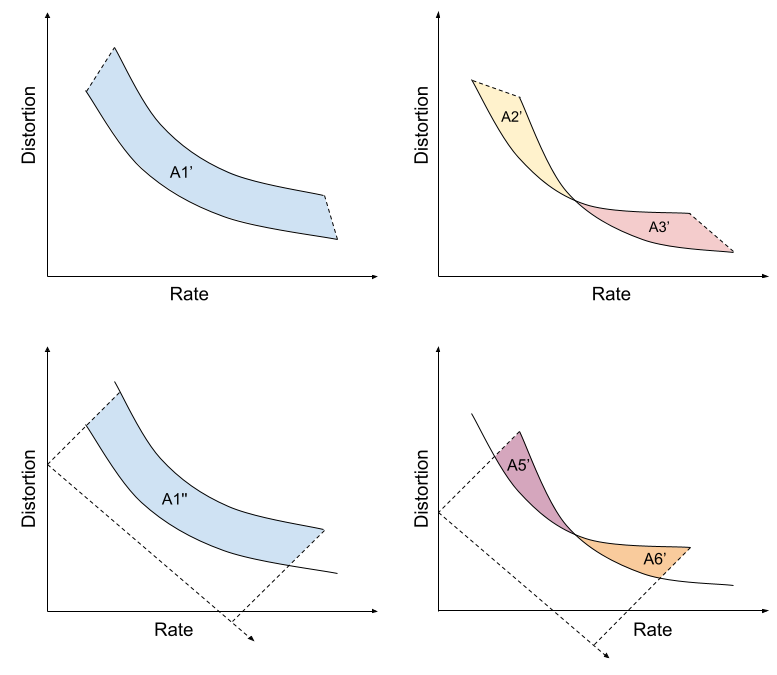}
        \caption{\ }
        \label{fig:bd2b}
    \end{subfigure}
    \begin{subfigure}{0.45\linewidth}
        \centering
        \includegraphics[trim={389px 338px 0 0},clip,width=\linewidth]{fig/queir_2.png}
        \caption{\ }
        \label{fig:bd2b}
    \end{subfigure}
    \begin{subfigure}{0.45\linewidth}
        \centering
        \includegraphics[trim={0 0 389px 338px},clip,width=\linewidth]{fig/queir_2.png}
        \caption{\ }
        \label{fig:bd2a}
    \end{subfigure}
    \begin{subfigure}{0.45\linewidth}
        \centering
        \includegraphics[trim={389px 0 0 338px},clip,width=\linewidth]{fig/queir_2.png}
        \caption{\ }
        \label{fig:bd2b}
    \end{subfigure}
    \caption{\ The alternative computation of BD metrics. In examples (a) and (b), the corresponding endpoints of the two RD curves  are connected, ensuring rotational invariance and independence from projection. Note that this alternative computation is not aligned with the conventional BD computation. In examples (c) and (d), the RD points are projected onto a line with inclination $\theta$ (slightly dislocated here for aesthetic purposes).}
    \label{fig:bd2}
\end{figure}

The BD metric has as input two sets of 2D points, say $(x_i,y_i)$ and $(u_i,v_i)$.
Each set is interpolated into $y(t)$ and $v(t)$, where $t$ is in a common scale for $x$ and $u$.
The BD is the average difference of their integrals over the region of interest. Hence, 
\begin{eqnarray}
\label{eq:bd}
\mbox{BD} & = & \Delta(\{x_i\}, \{y_i\}, \{u_i\}, \{v_i\}) \\ \nonumber
 & = & \frac{1}{t_1 - t_0} \int_{t_0}^{t_1} ( y(t) - v(t) )\ dt ,
\end{eqnarray}

\noindent where the limits of integration $t_0$ and $t_1$ are chosen from $(x_i,y_i)$ and $(u_i,v_i)$ in order to correspond to the overlapping region of the two RD curves under consideration along the rate (respectively, distortion) dimension when the BD-PSNR (respectively, BD-Rate) is computed. 

An important problem with BD is that it can lead to different conclusions in many cases. 
Consider the example in Fig. \ref{fig:bd1}, where in Fig. \ref{fig:bd1}(a) the BD-PSNR is illustrated as area $A_1$, while area $A_2$ in Fig. \ref{fig:bd1}(b) illustrates the respective BD-rate. 
The area ``under'' curves depend on the end points ($t_0$ and $t_1$ in Eq. (\ref{eq:bd})), which can be problematic. 
Consider the RD curves in Figs. \ref{fig:bd1}(c) and \ref{fig:bd1}(d), which cross themselves. 
BD-PSNR in Fig. \ref{fig:bd1}(c) may be positive if $A_4$ is larger than $A_3$, 
but BD-rate in Fig. \ref{fig:bd1}(d) may also be positive as $A_5$ is larger than $A_6$. These conclusions contradict with each other when it comes to decide which codec performs better in terms of their RD performance. 
An alternative is to change from quality to distortion measures, and compute the areas as connecting the end points of the RD curves
(Figs. \ref{fig:bd2}(a) and \ref{fig:bd2}(b)),
so that the areas are invariant to any rotation and do not depend on projection. However, this is not how BD is conventionally  computed. 

\subsection{BD generalization for projection orientation}
 
A codec is described by a number of points $(r_i,d_i)$ in the RD plane.
BD-MSE$^{dB}$ (instead of BD-PSNR) or $\Delta D$ is calculated by using distortion values as $y_i$ and $v_i$, while $x_i$ and $u_i$ are the respective rate values. 
BD-rate or $\Delta R$ is the contrary, i.e. using rate values as $y_i$ and $v_i$, while $x_i$ and $u_i$ are the respective distortion values. 

Assume one wants to jointly minimize rate and distortion, i.e. to minimize $J = D + \lambda R$ of a codec, and to maximize the difference $\delta J = \delta D + \lambda\ \delta R$ over an anchor codec. 
If one wants to minimize $J = D + \lambda R,$ one should get the point  $(r_i,d_i)$ that is the closest to a line of inclination $\theta$ where 
\[
\tan(\theta) = -\lambda. 
\]

If, instead of projecting the RD points onto the $R$ axis ($\theta=0$) in order to compute $\Delta D$ or onto the $D$ axis $(\theta=-\pi/2)$ in order to compute $\Delta R$, we project the points onto the line with inclination $\theta$ (Figs. \ref{fig:bd2}(c) and \ref{fig:bd2}(d)), then we can find the $(x_i,y_i)$ points by rotating the 
$(r_i,d_i)$ points as 
\[
x_i = r_i \cos(\theta) + d_i \sin(\theta) ,
\]
\[
y_i = -r_i \sin(\theta) + d_i \cos(\theta) .
\]

If we do the same with the $(r_i,d_i)$ points that are associated with $(u_i,v_i)$, then the $\Delta(\lambda)$ metric is computed by interpolating the points and evaluating the integral in Eq.~\eqref{eq:bd}. 
Note that if we let $\lambda = 0$ we are actually computing $\Delta(0) =\Delta D$, while if we let $\lambda = \infty$ we are actually computing $\Delta(\infty) = \Delta R$.
Intermediate values of $\lambda$ lead to differences in intermediate directions.

\subsection{BD metrics in RDC volumes}

$\Delta(\lambda)$ is applied on two functions of one dimension, thus having a unique domain. 
As we increase one additional dimension, we have multiple options to compute BD differences. 
Let us consider using the same $\Delta(\lambda)$ in the 3D case. 

\begin{figure}[tbp]
	\centering
        \includegraphics[width=0.8\linewidth]{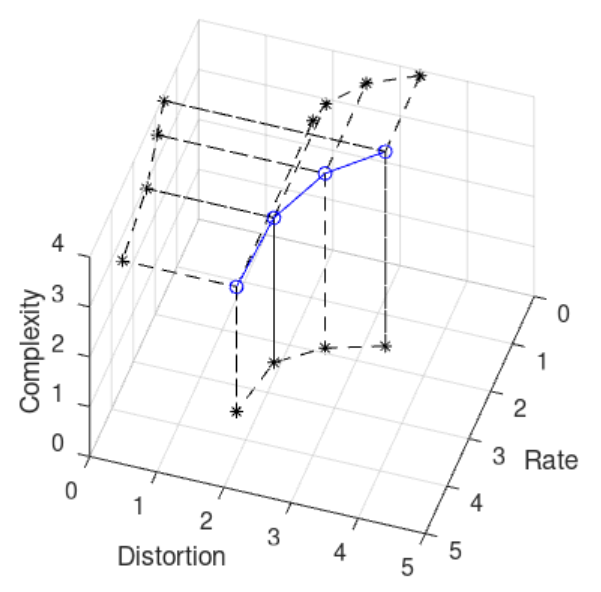}
        \caption{\ Toy example showing the operating points of a codec in RDC space, and the respective projections in the planes $R=0$, $D=0$, and $C=0$.}
	\label{fig:3dprojection2}
\end{figure}

Assume all the codecs under consideration are points in $(R,D,C)$ space. Each codec or coding instance is a point $(r_i, d_i, c_i)$.
All metrics discussed in the previous section can be obtained by zeroing $c_i$ and effectively projecting the point onto the $C=0$ plane (See Fig. \ref{fig:3dprojection2}). So, we use the $(r_i, d_i, 0)$ point coordinates to compute BD as discussed. We could then compute $\Delta R$, $\Delta D$, or intermediate $\Delta(\lambda)$ at will.  
Note that there is nothing special about using the $D$ axis along with the $R$ axis, and one could use the $C$ axis instead. We can also calculate $\Delta R$ if we project the points onto the $D=0$ plane, i.e. if we use $(r_i, 0, c_i)$ points. We could follow the same process and calculate $\Delta R$, $\Delta C$, or intermediate $\Delta(\lambda)$.
But also note that this new $\Delta R$ obtained using the $(R,0,C)$ plane is different from the $\Delta R$ obtained using the $(R,D,0)$ plane. So, we may call this new delta as $\Delta R_C$ and the other, from the previous section, as $\Delta R_D$. 
Similarly, we can call the delta on the $C$ axis as $\Delta C_R$ and the intermediate deltas as $\Delta_{RD}(\lambda)$ and $\Delta_{RC}(\lambda)$. 
In order to maintain a consistent notation, we can refer to $\Delta D$ using the $(R,D,0)$ plane as $\Delta D_R$. 
We can complement these metrics by projecting the points $(r_i, d_i, c_i)$ onto the $R=0$ plane, i.e. the $(0,D,C)$ plane. That yields the similar metrics $\Delta C_D$, $\Delta D_C$,  and $\Delta_{DC}(\lambda)$. 

In summary, in the RDC volume of points (see Fig. \ref{fig:3dprojection2}), we have  6 axis-oriented delta metrics: 
\[
\Delta R_D , \Delta R_C , \Delta D_R , \Delta D_C , \Delta C_R , \Delta C_D  ,
\]

\noindent where $\Delta R_D$ and  $\Delta D_R$ correspond to the traditional $\Delta R$ and  $\Delta D$ from the previous section.
There are also three intermediate metrics which can generate the above by using $\lambda =0$ or $\lambda = \infty$:  
\[
\Delta_{RD}(\lambda) ,  \Delta_{RC}(\lambda)  ,  \Delta_{DC}(\lambda) .
\]

Alternatively, we can define another 3D metric $\Delta_{RDC}(\lambda,\gamma)$ as the  integral of the difference of 2D functions, i.e. the volume in between two surfaces. There are two problems with that approach. 
Firstly, codecs are not usually represented as surfaces in RDC space. It is common to use quantizer parameters of quality factors such that the codec operation of a sequence is more typically represented as a curve in 3D space. 
Most neural codecs are not easily parameterized such that their coding instantiations would be clouds of 
$(r_i, d_i, c_i)$ points. 
The second problem is to define the direction to compute the differences, i.e. $(\lambda,\gamma)$. 
Of course, one can always project the functions onto one of three planes: RD, RC, or DC.
In the next section we propose a method to address these two concerns. 

\section{A cost in the RDC volume} 
\label{sec:costinRDC}

We can make a geometric illustration of the functions we are trying to minimize. 
In an $RDC$ volume, assuming non-negative $RDC$ values, and positive constants $\lambda$ and $\gamma$,  if one wants to minimize $J = D + \lambda R + \gamma C$ out of a cloud of points $\{p_n\} = \{(r_n, d_n, c_n)\}$, the one that has the lowest $J_{*n} = d_n + \lambda r_n + \gamma c_n$ is the one that is the closest to a {\em cost} plane 
\begin{equation}
	D + \lambda R + \gamma C = 0 . 
\end{equation}
The distance $z_i$ from point $p_i$ to the cost plane is: 
\begin{equation}
	z_i = \frac{d_i + \lambda\ r_i + \gamma\ c_i}{\sqrt{1 + \lambda^2 + \gamma^2}} . 
\end{equation}
The coordinates $(r_i^\prime, d_i^\prime, c_i^\prime)$ of the point where $p_i$ projects onto the plane are: 
\begin{eqnarray}
	d_i^\prime & = & d_i - q_i ,\\ 
	r_i^\prime & = & r_i - q_i \lambda , \\ 
	c_i^\prime & = & c_i - q_i \gamma , 
\end{eqnarray}

\noindent where 
\begin{equation}
	q_i = \frac{d_i + \lambda\ r_i + \gamma\ c_i}{1 + \lambda^2 + \gamma^2} . 
\end{equation}

\begin{figure}[tbp]
	\centering
	\includegraphics[width=0.635\linewidth]{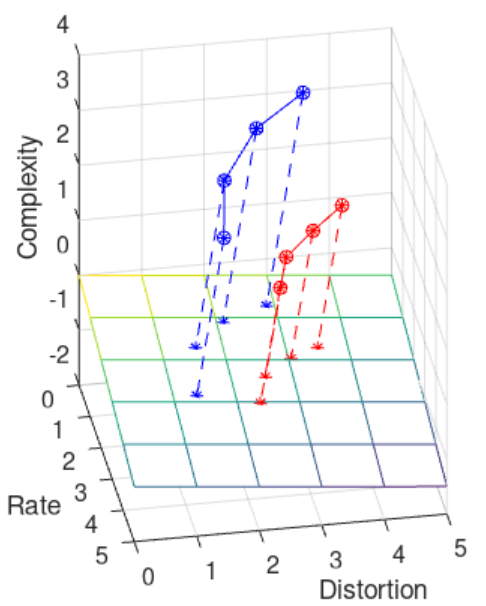}
	\caption{\ Projection of the RDC points of two hypothetical codecs onto the plane $D+2R+10C=0$. }
	\label{fig:3dprojection}
\end{figure}

A codec defined by a set of coding instantiations, or points in RDC space, is usually parameterized by, for example, a quantizer parameter (QP) forming a segmented curve of straight lines. 
The distance of each point to the plane is its cost. 
We want to measure the average distance of each curve to the plane, as illustrated in Fig. \ref{fig:3dprojection}. 
For every pair of adjacent points $p_i,p_j$ in this segmented curve of straight lines, the area forms a right trapezium. As such, the average distance of a segment in the curve to the plane is the average of the distances from $p_i,p_j$ to the plane.
Since we have segments with different lengths one may weight each segment average by its respective projected length. 
In symbols, we let the average distance (to the plane) of points $p_i$ and $p_j$ be
\begin{equation}
	z_{i,j} = \frac{z_i + z_j}{2}  , 
\end{equation}

\noindent 
and define the length of the projected segment as 
\begin{equation}
	\ell_{i,j} = \sqrt{(d_i^\prime-d_j^\prime)^2 + (r_i^\prime-r_j^\prime)^2 + (c_i^\prime-c_j^\prime)^2}  . 
\end{equation}

\noindent 
Then, the cost of the entire curve of $N$ points is

\begin{equation}
	\label{eq:cost_curve}
	J = \frac{\sum_{i=1}^{N-1} \ell_{i,i+1} z_{i,i+1}}{\sum_{i=1}^{N-1} \ell_{i,i+1}} . 
\end{equation}

\begin{figure}[tbp]
	\centering
        \includegraphics[width=0.65\linewidth]{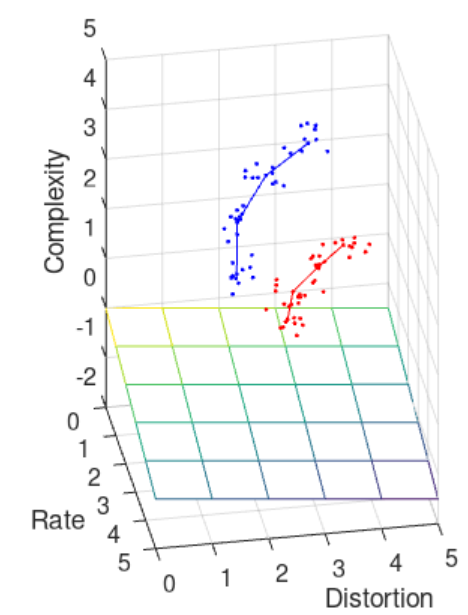}
	\caption{\ The RDC point clouds of two hypothetical codecs with different coding parameters over the plane $D+2R+10C=0$.}
	\label{fig:3dprojection3}
\end{figure}


The above equation is the preferred way to calculate the RDC cost of a codec that can be smoothly parameterized, for example using a QP. This is the case with non-neural video codecs and with some variable-rate neural video codecs.




However, there are cases where there may not exist such a parameter or there may be too many parameters such that we cannot assume a continuous RDC curve. In these cases, 
we may have a cloud of RDC points, as illustrated in Fig. \ref{fig:3dprojection3}.
Let there be multiple codecs to compare. 
The $m$-th codec yields multiple RDC points $\{p_{mn}\}$, each point with coordinates  ($r_{mn},d_{mn},c_{mn}$), which is then associated with the cost 
\begin{equation}
    J_{mn} = d_{mn} + \lambda  r_{mn} + \gamma c_{mn}.
\end{equation}
It is not obvious how to associate a cost with each codec based on the costs of its points. 
One could compute the minimum or mean, as examples. Let there be $N_r$ RDC points for each codec. We associate the minimum or mean cost with a codec as: 
\begin{equation}
	J_m = \min_{n} \{J_{mn}\} \ \ \ \  \mbox{or} \ \ \ \ J_m = \frac{1}{N_r} \sum_n  J_{mn}.
\end{equation}

In summary, we can calculate the RDC cost as in Eq. \eqref{eq:cost_curve} for a variable-rate codec, or as above for codecs with individual RDC points.

\section{Application space}
\label{sec:applicationexample}

In Eq.(\ref{eq:linearcost}), the minimization of $J$ ensures operating at the $K$-dimensional lower convex hull of all cost points $s_i$, i.e. at a hypersurface parameterized by all $\lambda_i$. 
If we, however, have a finite number of codec candidates, how do we compare them? 
In order to calculate $J$, we need to set vector $\bf u$. 
Back to our reduced case of interest, in order to compare codecs we need to set $\lambda$ and $\gamma$.
How do we choose such parameters? 
The answer depends on the application. 

A problem in comparing, for example, two codecs in an RDC space, is to conciliate the different quantities that are combined. 
Rate is given in bits per second, distortion is given in whatever the unit we use to measure the pixel brightness, while complexity, as here defined, is measured in kMAC/pixel (or other metric). 
The result of the summation of such diverse quantities may be difficult to understand.

We propose to define an application and to map each RDC quantity to a common cost for that application. 
For simplicity, we can make a linear map from the RDC cost metrics to a common cost:  

\begin{equation}
\alpha_1 D + \beta_1 + \alpha_2 R + \beta_2 + \alpha_3 C + \beta_3 . 
\end{equation}

\noindent 
As the constant terms do not affect minimization, this is the same as minimizing

\begin{equation}
J^\prime = \alpha_1 D + \alpha_2 R + \alpha_3 C. 
\end{equation}

For example, we can set $J^\prime$ to be the global monetary cost for operating the codec. 
We need to find the $\{\alpha_i\}$ that make the proper translation from RDC units to their contribution to the cost for operating the codec. 

Consider the example of a hypothetical streaming application: there are 1000 users to transmit and decode 1000 hours of HD, 30Hz, video. 
Also, consider the following issues:

\begin{itemize}
	\item As $D$ increases, customers avoid using the codec. Then $\alpha_1$ should reflect the cost associated with the loss of customers when choosing that specific codec, or the inverse of the revenue of the number of users in the system.  
	\item As $R$ increases, one needs to rent more bandwidth. Then, $\alpha_2$ should reflect the cost of renting the necessary bandwidth infrastructure. 
	\item As $C$ increases, it becomes more expensive to build and operate the codec, consuming more energy. Then, $\alpha_3$ should reflect the fixed and variable costs of operating the codec.   
\end{itemize}

We have the following assumptions: 

\begin{itemize}
	\item Users pay twice the video transmission costs to cover other costs such as salaries, taxes and profits. 
	\item Reference quality is to use H.265/HEVC operating at a rate of 3 Mb/s.  
	\item Costs are: hardware, energy, and bandwidth.  
	\item The average energy price in the US is US\$ 0.15/KWh.
	\item The average cost of transferring data is US\$ 0.08/GB, based on some Microsoft Azure\textsuperscript{\textregistered} charges. 
	\item An NVIDIA RTX 3080 is capable of processing 4K, 30Hz video at 128 kMAC/pixel, draws 320W of power, and costs around US\$ 700.
\end{itemize}

From the above assumptions, one may say the NVIDIA RTX 3080 has a fixed cost of $700/128 = $ US\$ 5.46 /(kMAC/pixel), and that it consumes, in an hour, 0.32 KWh at a cost of US\$ 0.048, yielding an expense rate of
$0.048/128 = $ US\$ 0.000375 / (kMAC/pixel) / hour. 
One kMAC/pixel for 1000 hours of video decoding would demand a cost of $5.46 +1000 \times 0.000375 = 5.843$ US\$/(kMAC/pixel), such that
one thousand users would spend US\$ 5,843 / (kMAC/pixel). 


For every Mb/s of rate, sending 1000 hours of video to 1000 users consumes 450,000 GB of data at a cost of US\$ 36,000. 
Then, the rate costs US\$ 36,000 for every increase of 1 Mb/s. 

Assume the reference H.265/HEVC demands as much complexity as 7 kMAC/pixel and compresses video at 3 Mb/s. 
Hence, operating costs would be 7$\times$5,843 + 3$\times$36,000 = US\$ 148,901. 
When we double this number to cover the other costs, we get US\$ 297,802, which we could round to US\$ 300K of revenue. 
If we wild guess that at 40 dB PSNR quality (0-255, 8-bit images) all 1000 users are subscribers, but at 30 dB all users would leave, it then translates into a loss of approximately US\$ 300,000 as MSE jumps from 6.501 to 65.01, i.e. an increase in MSE of 58.51.  
This is the same as having a cost of US\$ 5,127 per unit increase in MSE.


If we measure $D$ in (8-bit) MSE, $R$ in Mb/s and $C$ in kMAC/pixel, an increase of one unit in $D$ generates a cost of US\$ 5,127, 
an increase of one unit in $R$ generates a cost of US\$ 36,000, and an increase of one unit in $C$ would cost US\$ 5,843. 
Hence, 
\[\alpha_1 = \textup{5,127}, \ \ \ \alpha_2 = \textup{36,000}, \ \ \ \alpha_3 = \textup{5,843}, \  \]

\noindent such that if we want to minimize the monetary cost of the operation we should use: 
\begin{equation}
	J^\prime =  \textup{5,127}\ D + \textup{36,000}\ R + \textup{5,843}\ C . 
\end{equation}

If one wants a cost function $J$ to reflect such a monetary cost $J^\prime$, we can simply  normalize the weights $\lambda$ and $\gamma$ by dividing all numbers by $\alpha_1$, such that 
\[
\lambda \approx \textup{7.02} \ \ \ \mbox{and} \ \ \ \gamma \approx \textup{1.14}\ . 
\]

Clearly, the calculation of all $\{\alpha_i\}$, hence of $(\lambda,\gamma)$, has to be customized for each application.
Once $(\lambda,\gamma)$ are set, any two codecs can be compared in an RDC sense for the given application. 
Thus, every application is associated to a $(\lambda,\gamma)$ pair, spanning what we refer as the  application space.

In summary, the proposed metric is as follows. Given an application, one maps the RDC costs to a single cost (monetary or otherwise), thus calculating the $\{\alpha_i\}$. 
With the $\{\alpha_i\}$ at hand, one calculates the $(\lambda,\gamma)$ point where the given application lies. With the $(\lambda,\gamma)$ pair, and an RDC point, one can calculate the overall cost $J$.
Two video codecs, after encoding a test set with their intrinsic settings, can be directly compared, and we should pick the one with the lowest $J$. 

\section{codec-centric analysis}
\label{sec:coder-centric}

Can we analyze a codec without choosing $\lambda$ and $\gamma$?
We can do a codec-centric analysis, wherein we find the application-space regions, i.e. ($\lambda$,$\gamma$) ranges, for which a codec is advantageous or has a lower cost. 
For this we can choose a number of ($\lambda$,$\gamma$) points, calculate their costs $J$ and interpolate the rest. 
The result is an application plane $J(\lambda,\gamma)$ which indicates the best-suited applications for the given codec. 

Yet another alternative is to examine a collection of available codecs, and to decide which codec is the best for each point in application space. 
Let there be $N_c$ codecs to test, each instantiated at $N_r$ different rates. 
The $j$-th rate of the $i$-th codec yields points $P_{ij}$ located at $(R_{ij}, D_{ij}, C_{ij})$ in RDC space. 
We can then guess many values of $\lambda$ and $\gamma$, calculate their costs $J_{ij}$ as 
\begin{equation}
\label{eq:cost}
	J_{ij} = D_{ij} + \lambda R_{ij} + \gamma C_{ij}  ,
\end{equation}

\noindent 
and interpolate the values for the rest of the $J_{ij}(\lambda,\gamma)$. 

From the discussion in Sec. \ref{sec:costinRDC}, for each $(\lambda,\gamma)$, we can compute the cost $J_i$ of each codec as 

\begin{equation}
\label{eq:mincost}
J_i(\lambda,\gamma) = \min_j J_{ij}(\lambda,\gamma)  
\end{equation}

\noindent or 

\begin{equation}
\label{eq:averagecost}
	J_i(\lambda,\gamma) = \frac{1}{N_r}\sum_{j} J_{ij}(\lambda,\gamma) . 
\end{equation}
At this point we would have $N_c$ surfaces $J_i(\lambda,\gamma)$, one for each codec, covering the whole application space. 
The function $J_i(\lambda,\gamma)$ is usually non-uniformly increasing with both $\lambda$ and $\gamma$.
The relevant information comes from comparing two or more codecs. 
One can select a reference codec and compare the cost functions of one against the other, thus revealing the regions in the $(\lambda,\gamma)$ space where each codec is best suited. 
Another useful information that emerges from these maps is to compare them all and pick the best codec for a given point in the $(\lambda,\gamma)$ space, i.e.
\begin{equation}
	B(\lambda,\gamma) = {\rm arg} \min_i J_{i}(\lambda,\gamma) 
\end{equation}

\noindent indicates the best codec, among all candidates, for that application point. 
A profound result of examining $B$ is that only the codecs whose indices are in the $B$ map would have any sense pursuing for those given metrics of R, D, and C. 
The other codecs may only make sense with different complexity metrics or using a different cost vector $\bf c$. We remark that our analysis framework does not impose any constraint on the complexity metrics to be used.

\section{Experimental results}


\begin{table}[tbp]
\centering
\caption{Evaluated codecs.}
\label{tab:codecs}
\begin{tabular}{ccc}
\hline
\multirow{2}{*}{$i$} & \multirow{2}{*}{codec} & Decoding complexity \\
& & (kMACs/pixel) \\ \hline
0   & CANF-VC \cite{CANF_VC} & 1748                              \\
1   & DCVC \cite{DCVC}        & 762                               \\
2   & DCVC-TCM \cite{DCVC_TCM}    & 924                               \\
3   & DCVC-HEM \cite{DCVC_HEM}    & 1252                              \\
4   & DCVC-DC \cite{DCVC_DC}     & 924                               \\
5   & DCVC-FM \cite{DCVC_FM}     & 878                               \\
6   & MaskCRT \cite{MASKCRT}  & 767                               \\
7   & C16 \cite{MMSP}         & 541                               \\
8   & C32 \cite{MMSP}         & 592                               \\
9   & C64 \cite{MMSP}         & 762                               \\
10  & CR16 \cite{MMSP}        & 541                               \\
11  & CR32 \cite{MMSP}        & 593                               \\
12  & CR64 \cite{MMSP}        & 764                               \\
13  & MCR16 \cite{MMSP}       & 598                               \\
14  & MCR32 \cite{MMSP}       & 649                               \\
15  & MCR64 \cite{MMSP}       & 821                               \\
16  & HyTIP \cite{HyTIP}       & 873                              
\end{tabular}
\vspace{-1em}
\end{table}


We have analyzed data comprised of four RDC points of 17 neural video codecs, including several state-of-the-art codecs and mainstream coding frameworks that are based on conditional coding, conditional residual coding, and masked conditional residual coding, as shown in Table \ref{tab:codecs}. More details about Cnn , CRnn , MCRnn, and HyTIP codecs in Table~\ref{tab:codecs} are provided in the appendix, which is submitted as a separate document.
Each of these 16 neural video codecs (except DCVC-HEM, DCVC-DC, and DCVC-FM) is trained to support four different rates with 4 distinct sets of network weights. Unlike the other neural video codecs, DCVC-HEM, DCVC-DC, and DCVC-FM are variable-rate codecs; they are able to compress an input video for a continuous range of bit rates. For these codecs, we produce 4 distinct bit rates corresponding to 4 quality levels.
There are, thus, 64 coding instantiations to appreciate, generating 64 $(r_{ij},d_{ij},c_{ij})$ points, from which the 16 costs $J_i$ for the respective 16 codecs are computed according to Eq. \eqref{eq:mincost} or \eqref{eq:averagecost} .

\begin{table}[tbp]
\centering
\caption{Test datasets.}
\label{tab:datasets}
\begin{tabular}{ccc}
\hline
Dataset & Number of sequences & Resolution \\
\hline
UVG \cite{UVG}  & 6 & 1920$\times$1080 \\
HEVC-B \cite{HEVC_Dataset} & 5 & 1920$\times$1080 \\
HEVC-C \cite{HEVC_Dataset} & 4 & 832$\times$480 \\
HEVC-E \cite{HEVC_Dataset} & 3 & 1280$\times$720 \\
HEVC-RGB & 6 & 1920$\times$1080 \\
MCL-JVC \cite{MCL-JVC} & 30 & 1920$\times$1080 \\  
\end{tabular}
\vspace{-1em}
\end{table}


The tests were carried to  compress all the sequences in the test sets
indicated in Table \ref{tab:datasets}.
There are 55 sequences, each with a different bit-depth, resolution, frame-rate, and number of frames.
For each sequence and codec, rate is calculated in megabits per second (Mb/s). 
In order to average $R$ across codecs, each rate value is normalized by converting it to bits/pixel and then denormalizing for a $1920\times 1080$-pixel 30 Hz sequence. 
It is worth noting that rate $R$ for each codec and configuration is computed as the average of all the 55 normalized rates, in Mb/s.
Each of the 64 coding instantiations demand encoding all the testsets.  
Distortion $D$ for each codec is computed as MSE (8-bit/color/pixel, i.e. 0-255 range) in the RGB domain, since not all neural video codecs support YUV coding. 
For 12-bit/color/pixel sequences, MSE is divided by 256 to make it compatible with the MSE of 8-bit sequences.  
Finally, $D$ for a given codec and configuration is computed by averaging all the 55 distortions in MSE for each sequence. 
To quantify the complexity $C$ of these neural codecs, we chose the decoding complexity in kMACs/pixel, which is intrinsic to their decoder architectures, as shown in Table \ref{tab:codecs}. Note that similar evaluations can be made using the encoding complexity. 

\begin{figure}[tbp]
    \centering
    \includegraphics[width=1.0\linewidth]{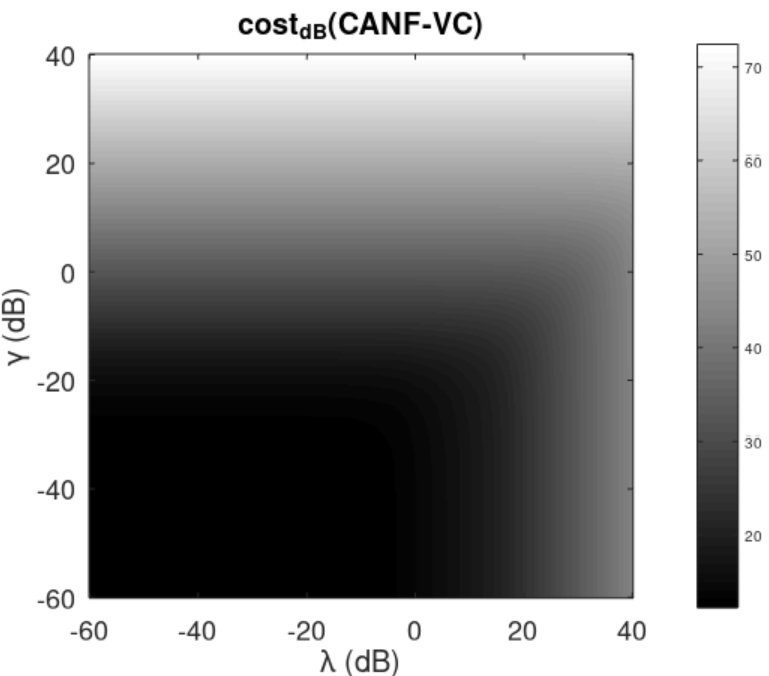} 
    
    \includegraphics[width=1.0\linewidth]{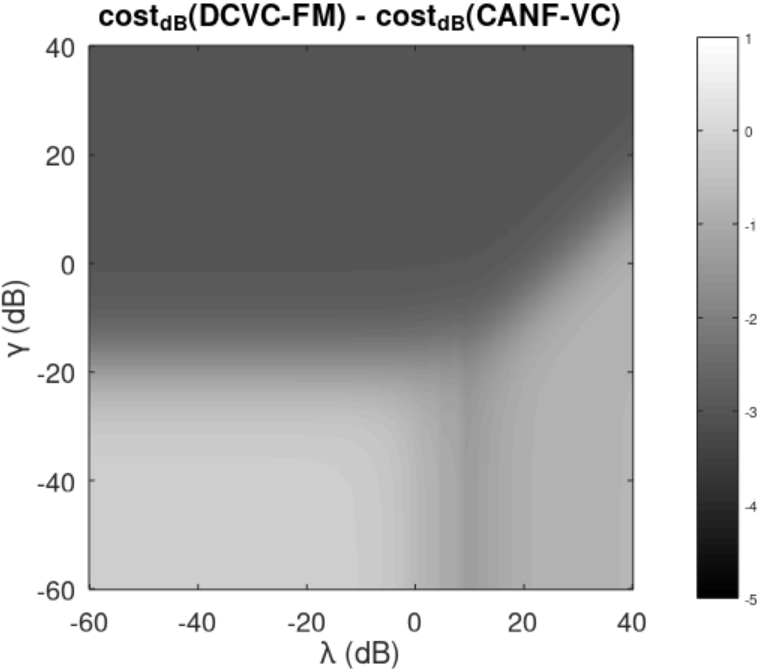} 

	\caption{\  Cost in application space $J_i(\lambda,\gamma)$, in dB, for CANF-VC, and the differences between the costs of DCVC-FM and CANF-VC.}
	\label{fig:3dsurf_appspace1}
\end{figure}

\begin{figure}[tbp]
    \centering
    \includegraphics[width=1.0\linewidth]{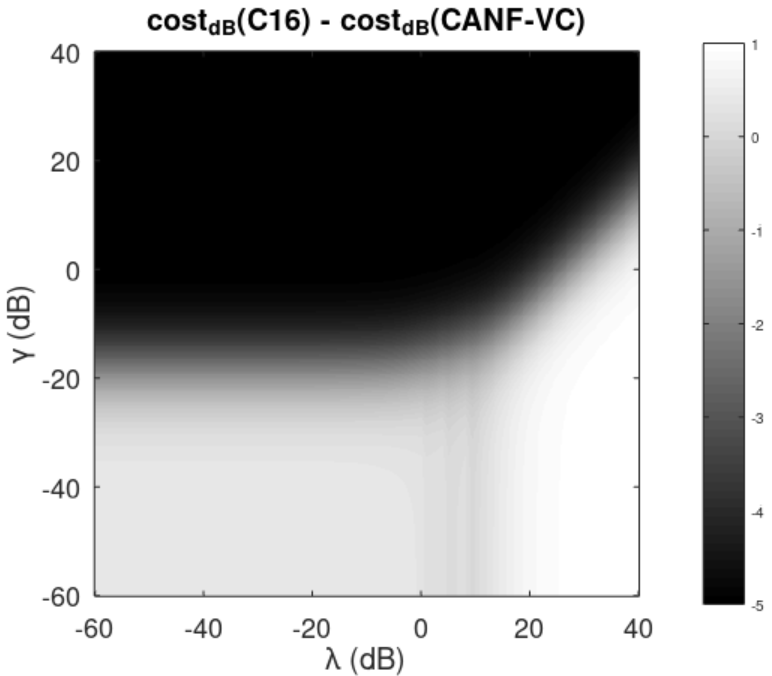} 

    \includegraphics[width=1.0\linewidth]{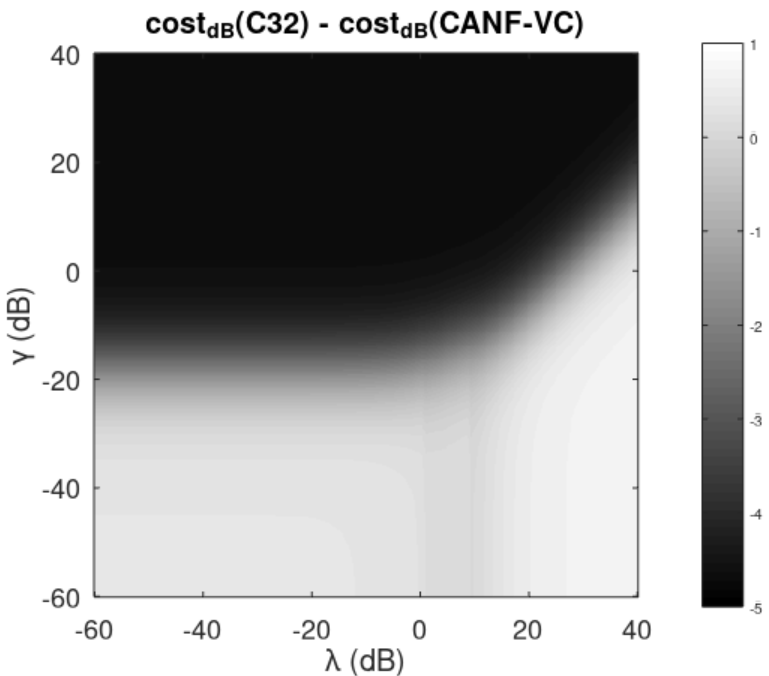}

	\caption{\ Cost differences in application space $J_i(\lambda,\gamma)$, in dB, between C16, C32 and CANF-VC.} 
	\label{fig:3dsurf_appspace2}
\end{figure}

For the $i$-th coder, we computed its cost in application space $J_i(\lambda,\gamma)$ as illustrated in the top of 
Fig. \ref{fig:3dsurf_appspace1} for an example codec, the CANF-VC. 
It is a monotonic function in $\lambda,\gamma$ as expected, and all axes are given in dB, i.e. $10\log_{10}(\lambda)$, $10\log_{10}(\gamma)$
 and $10\log_{10}(J)$.
When we compare codecs, we can obtain more information by subtracting their cost functions. 
This is illustrated in the bottom of Fig. \ref{fig:3dsurf_appspace1}, where the differences between the cost functions of DCVC-FM and CANF-VC are evaluated. 
In the bottom of Fig. \ref{fig:3dsurf_appspace1}, DCVC-DC improves over CANF-VC when it yields less cost, i.e. for negative (darker) regions. 
This holds true especially in the regions with higher $\gamma$ and/or $\lambda$ values (see Eq. \eqref{eq:cost}), which tells that DCVC-DC is less complex and performs better at lower rates. 
Hence, DCVC-DC would be preferred over CANF-VC for applications that demand more compression and are less tolerant to complex decoders.  
Similar results are shown in Fig. \ref{fig:3dsurf_appspace2}, where C16 and C32 are compared against CANF-VC. 
We stress that these observations are valid under our current RDC measures. They may vary according to the specification of RDC measures, which depends highly on application requirements.

\begin{figure}[tbp]
	\centering
	\includegraphics[width=0.99\linewidth]{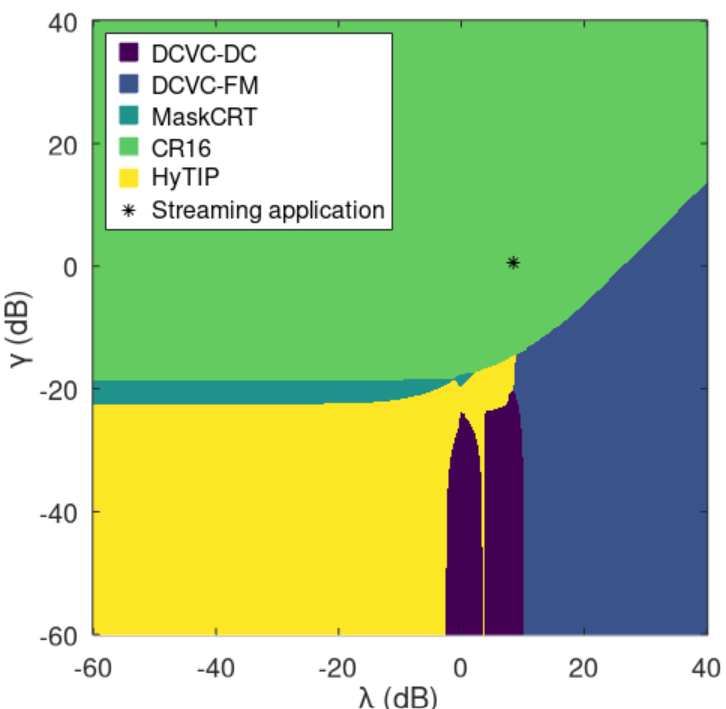} 
	\caption{\ Map of best codecs in $B(\lambda, \gamma)$ space considering the minimum value - Eq. (\ref{eq:mincost}).}
	\label{fig:appspace_coders_min}
\end{figure}

\begin{figure}[tbp]
	\centering
	\includegraphics[width=0.99\linewidth]{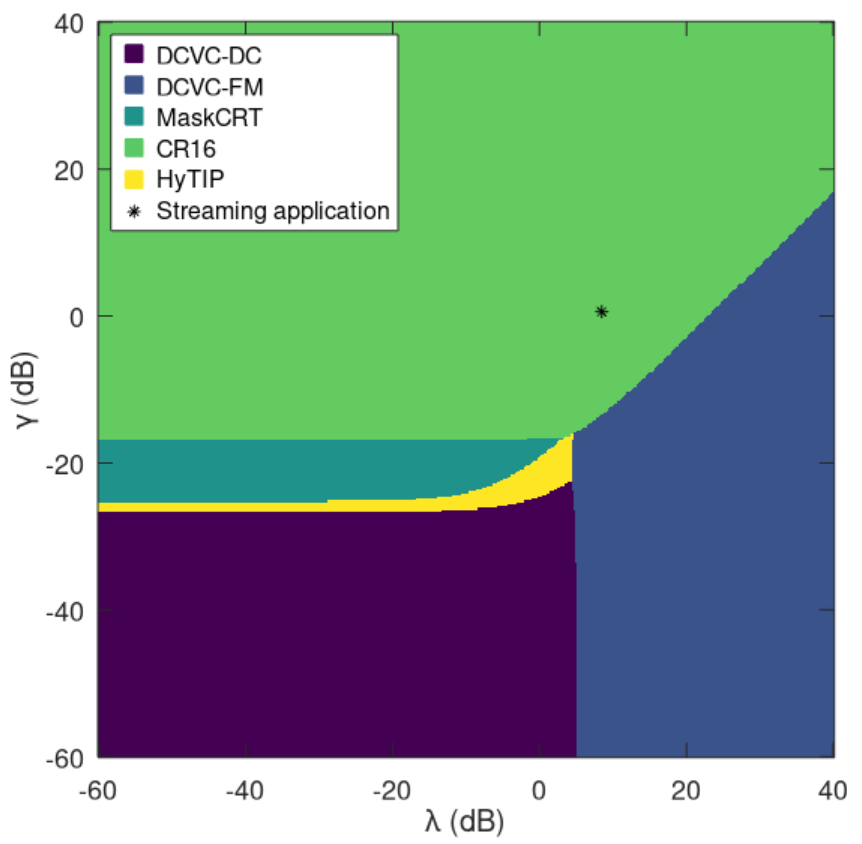} 
	\caption{\ Map of best codecs in $B(\lambda, \gamma)$ space considering the average value - Eq. (\ref{eq:averagecost}).}
	\label{fig:appspace_coders_mean}
\end{figure}

The previous comparisons were conducted for pairs of codecs. 
The maps $B(\lambda,\gamma)$ showing the best of all of the 16 codecs are given in Figs. 
\ref{fig:appspace_coders_min} and \ref{fig:appspace_coders_mean}, where we used Eqs. \eqref{eq:mincost} and \eqref{eq:averagecost} to evaluate $J_i(\lambda,\gamma)$, respectively. 
In both figures, we also include the application point associated with our ``toy'' streaming application in Sec. \ref{sec:applicationexample}. 

From Figs. \ref{fig:appspace_coders_min} and \ref{fig:appspace_coders_mean}, we learn that the best codec for our {\em toy} application is CR16. 
CR16 dominates the application space, and for applications with similar $(\lambda,\gamma)$ to ours, we expect CR16 to be the best as well. 
DCVC-DC, DCVC-FM, MaskCRT and HyTIP are also candidates for best codecs for some other applications that lie in some other regions of the $(\lambda,\gamma)$ space. 
These observations are conditioned on our RDC measures, i.e. $R$ in Mb/s, $D$ in MSE, and $C$ measured as the decoder's kMAC/pixel. 
The use of other RDC measures may lead to different results.

\begin{figure}[tbp]
	\centering
	\includegraphics[width=0.97\linewidth]{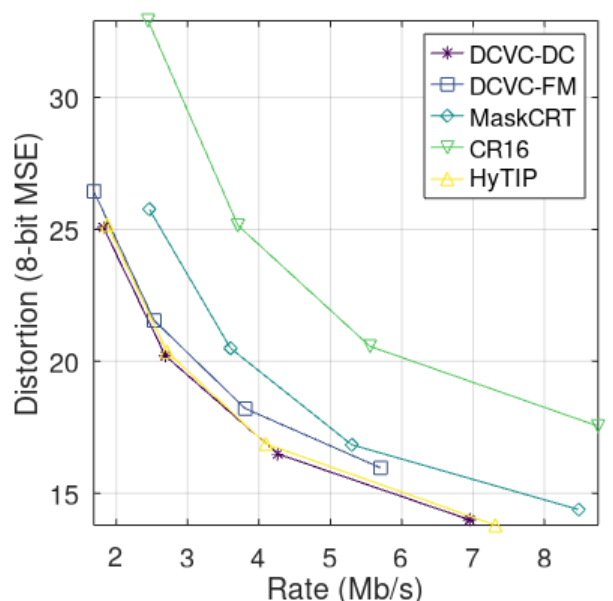} 
	\caption{Rate-distortion points for the winning codecs shown in Fig. \ref{fig:appspace_coders_min}.}
	\label{fig:gamma0_analysis1}
\end{figure}

\begin{figure}[tbp]   
    \includegraphics[width=0.97\linewidth]{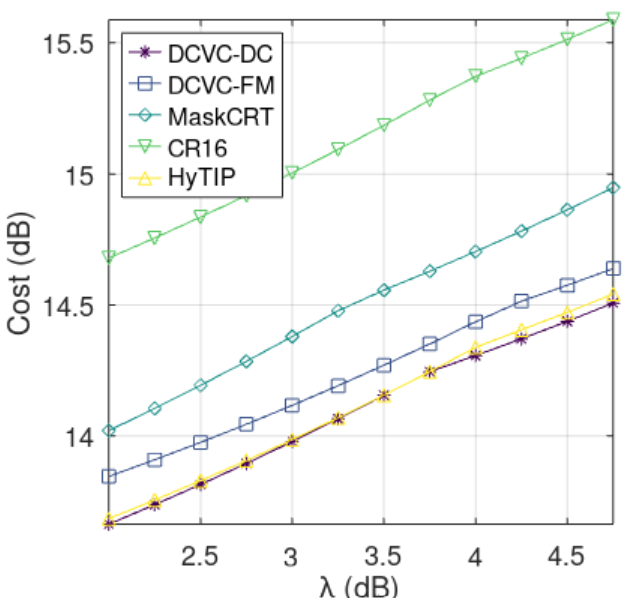} 
	\caption{$J_i(\lambda,0)$ values  for the winning codecs shown in Fig. \ref{fig:appspace_coders_min},  }
	\label{fig:gamma0_analysis2}
\end{figure}



Note that CR16 is the top performer for higher $\gamma$ values, namely, for applications where higher complexity implies a higher penalty. 
In the regions with lower $\gamma$ values, there are sudden transitions in Fig. \ref{fig:appspace_coders_min}. 
To understand this point, consider the extreme case where $\gamma = 0$, which reduces to the traditional RD analysis. 
In Fig. \ref{fig:gamma0_analysis1}, we show the RD points (the projections of the RDC points onto the $C=0$ plane) that were used to compute the costs $J_i(\lambda,0)$ with Eq. \eqref{eq:mincost} relative to the best codecs in Fig. \ref{fig:appspace_coders_min}.
Note how the points in the lower-convex hull alternate between DCVC-DC and HyTIP. 
If one computes the costs for such data, one obtains the $J_i(\lambda,0)$ curves, which are visualized and compared for a particular range of $\lambda$ in Fig. \ref{fig:gamma0_analysis2}. 
The result 
shows the alternation that leads to the patterns at the bottom of Fig. \ref{fig:appspace_coders_min}. 

\clearpage

\section{Conclusions}

We were motivated by the needs to analyze the complexity of video codecs, particularly with the rapid development of neural video codecs. 
We recognize that the complexity analysis of a codec is a multifaceted problem. There may be various costs to consider. Moreover, these costs may be weighted differently according to the application requirements. We propose using the linear Lagrangian combination of rate (R), distortion (D) and complexity (C). In the process, we addressed a few issues with the BD metrics and extended the metrics to an RDC volume. 

We developed the concept of the application space, where the points in the application space are the Lagrangian weights for the RDC variables. Specifically, we translate the application requirement for a codec into one of these points. We then convert the RDC values of the codec into a single variable reflecting, for example, the monetary cost of operating the codec. 
This novel approach allows for an objective analysis of various codecs.  
We demonstrated its utility in a toy streaming application. 
A total of 17 neural video codecs are analyzed and compared. The results showed that among all the codecs analyzed, only DCVC-DC, DCVC-FM, MaskCRT, CR16 and HyTIP have a chance of being the best codecs under our RDC measures. In particular, for the streaming application we proposed, the CR16 codec is the best choice.
Our approach can accommodate various RDC measures, being generalizable and useful for analyzing and adopting codec proposals in standards committees or industry.

\section*{Appendix}

\begin{figure}[h!]
	\centering
	\includegraphics[width=0.94\linewidth]{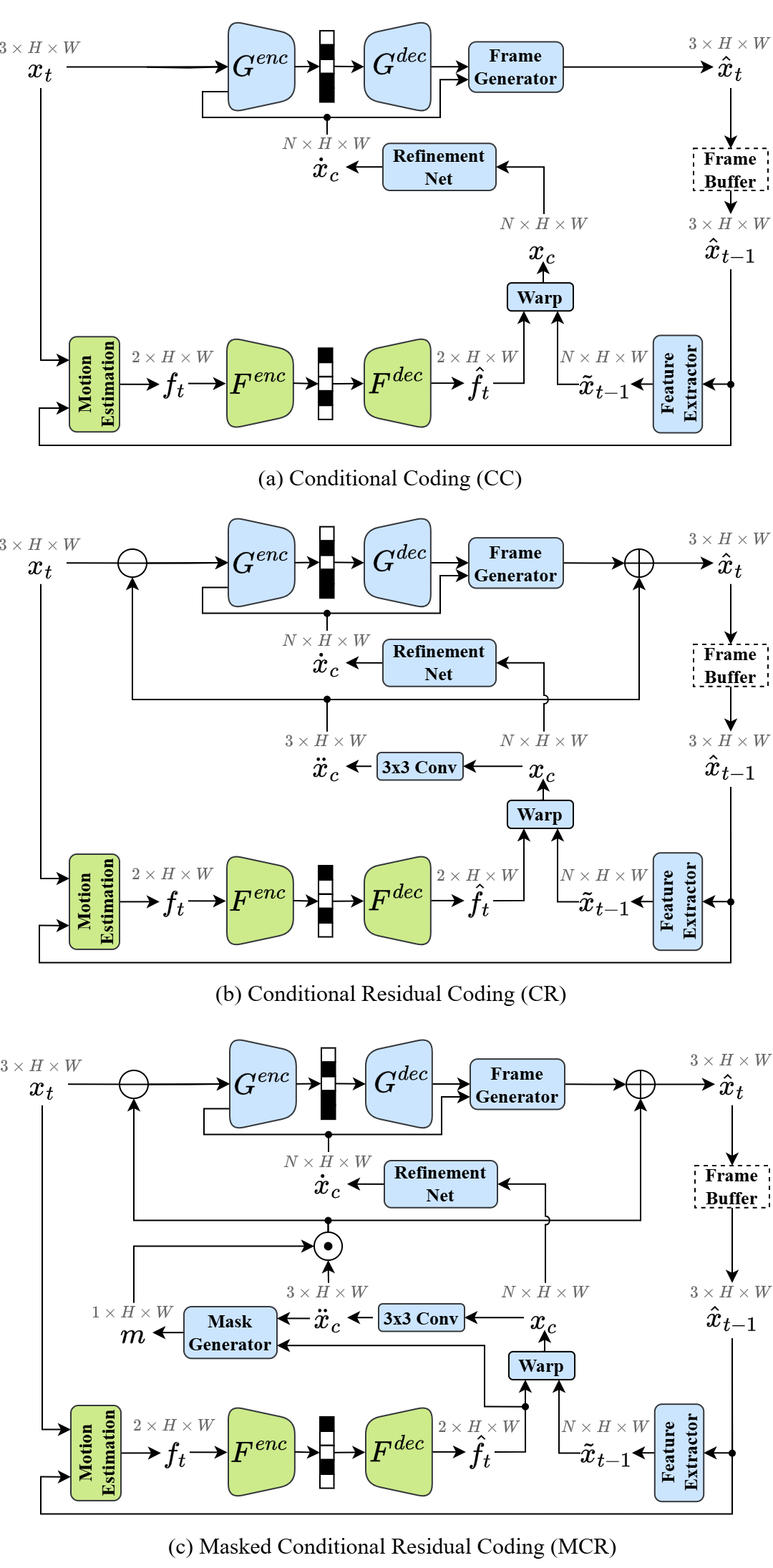}
	\caption{Illustration of (a) conditional coding, (b) conditional residual coding, and (c) masked conditional residual coding in [28]. Adapted from [28].}
	\label{fig:MMSP}
\end{figure}

To assist with the understanding of our paper, Fig.~\ref{fig:MMSP} illustrates the conditional coding (CC), conditional residual coding (CR), and masked conditional residual coding (MCR) from [28]. These frameworks are built upon the same code base and share nearly the same coding components. As shown, the major differences between CR and CC are that (1) CR introduces a $3\times3$ convolutional layer to generate a pixel-domain predictor $\ddot{x_c}$, and that (2) CR encodes the residue $x_t-\ddot{x_c}$ between a coding frame $x_t$ and its pixel-domain predictor $\ddot{x_c}$, conditioned on the temporal predictor $\dot{x}_c$. In comparison, CC directly encodes $x_t$ based on $x_c$. On top of CR, MCR further introduces a mask generator to produce a pixel-domain soft mask $m$, and blends $x_t$ and $x_t - \ddot{x}_c$ as $(1-m) \odot x_t + m \odot (x_t - \ddot{x}_c)$, which amounts to $x_t-m \odot \ddot{x}_c $ with $\odot$ denoting element-wise multiplication. In [28], the trade-offs between complexity and coding performance are explored by varying the channel size $N$ of $\tilde{x}_{t-1}$, $x_c$, and $\dot{x}_c$. This work uses C16/C32/C64 [28] to refer to the CC with $N=16, 32, 64$, respectively. Likewise, we use CR16/CR32/CR64 for the variants of CR and MCR16/MCR32/MCR64 for the variants of MCR. 

\begin{figure*}[h!]
	\centering
	\includegraphics[width=1\linewidth]{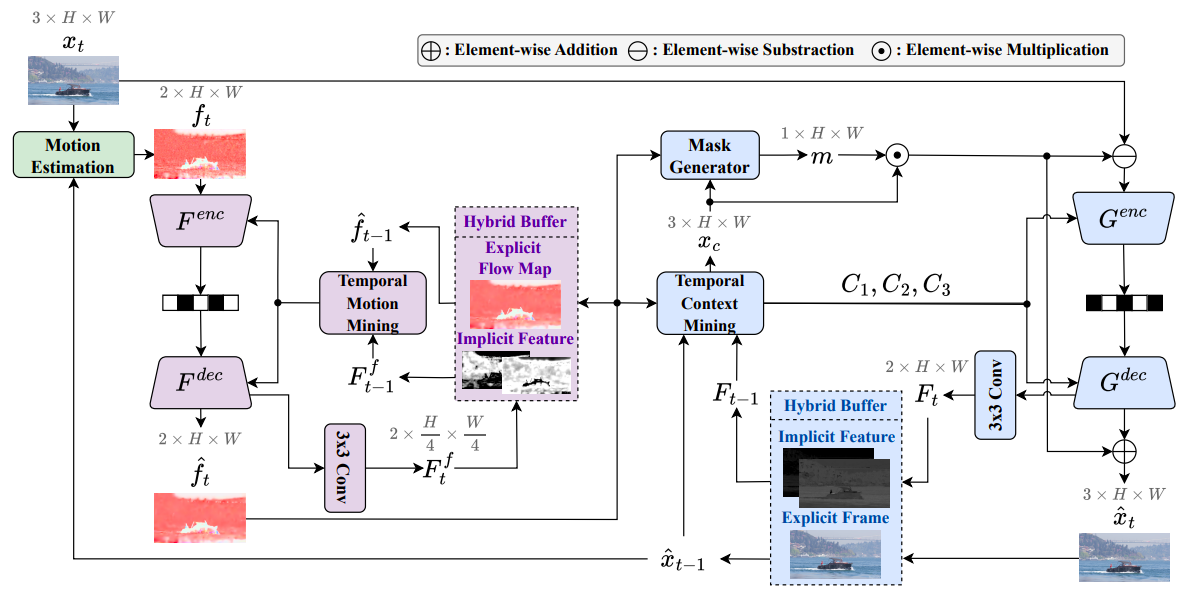}
	\caption{Illustration of HyTIP compression framework. Source: [29].}
	\label{fig:hytip}
\end{figure*}

HyTIP [29], illustrated in Fig.~\ref{fig:hytip}, is a compression framework built on the MCR architecture that modifies the temporal propagation path by introducing a hybrid buffer. In neural video compression, temporal information can be propagated through either an implicit buffer, which stores high-resolution latent features from the previous frame’s decoding ($F_{t-1}$) to guide the next frame’s coding, or an explicit buffer, which stores the previously decoded frame in the pixel domain ($\hat{x}_{t-1}$). HyTIP combines both approaches: at each coding step, the explicit buffer provides $\hat{x}_{t-1}$, which is used to generate the optical flow map $f_t$, while the implicit buffer supplies $F_{t-1}$. These two sources of temporal information are jointly used as condition signals to generate both the pixel-domain predictor $x_c$ and the multi-scale feature-domain predictors $\{C_1, C_2, C_3\}$. The same hybrid buffering strategy is also applied to motion coding, where both the decoded optical flow map $\hat{f}{t-1}$ (explicit) and the motion-latent features $F^{f}_{t-1}$ (implicit) are stored.

\bibliographystyle{IEEEtran}
\bibliography{references}

\begin{IEEEbiography} [{\includegraphics[width=1in,height=1.25in,clip,keepaspectratio]{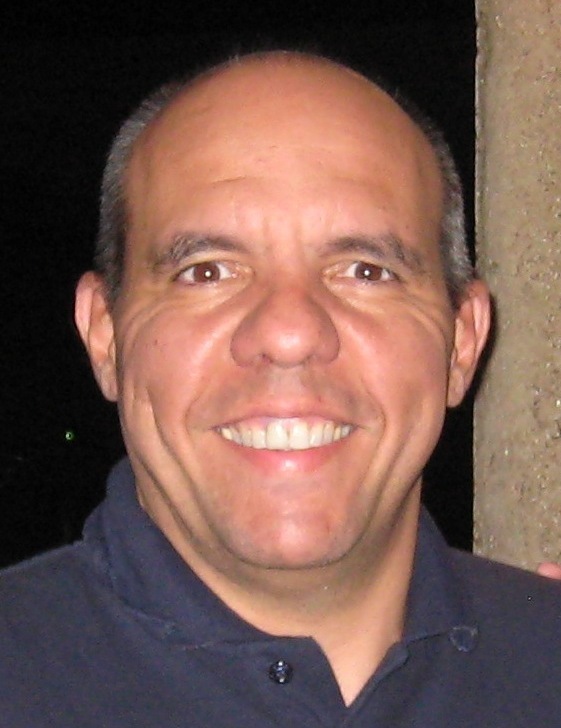}}]{Ricardo L. de Queiroz} 
(F'17) received the Engineer degree from Universidade de Brasilia , Brazil, in 1987, the M.Sc. degree from Universidade Estadual de Campinas, Brazil, in 1990, and the Ph.D. degree from The University of Texas at Arlington , in 1994, all in Electrical Engineering.
Since 2010, he has been a Full (Titular) Professor at the Computer Science Department at Universidade de Brasilia. 

Dr. de Queiroz has published extensively in Journals and conferences and contributed chapters to books as well. He also holds 51 issued patents. He has been a member of many Technical Committees and Editorial Boards of IEEE Societies. He has also been appointed an IEEE Signal Processing Society Distinguished Lecturer. He was Chair or part of the organizing committee of many flagship conferences of the Signal Processing and Circuits and Systems Societies of IEEE. His research interests include image and video compression, point cloud compression, multirate signal processing, and color imaging. Dr. de Queiroz is a Fellow of IEEE, CNPq Researcher level 1A, and a Senior Member of the Brazilian Telecommunications Society. 
\end{IEEEbiography}
\vspace{-3em}

\begin{IEEEbiography}[{\includegraphics[width=1in,height=1.25in,clip,keepaspectratio]{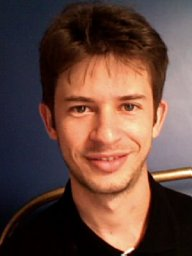}}]{Diogo C. Garcia} received the B.S., M.Sc., and Ph.D. degrees in Electrical Engineering from the University of Brasilia, Brazil, in 2006, 2008, and 2012, respectively. He is currently a Faculty Member with the Faculty of Engineering Sciences and Technology, University of Brasilia, and a Senior Member of the IEEE. His current research interests include image and video coding, super resolution, and multi-view and 3D processing.
\end{IEEEbiography}
\vspace{-3em}

\begin{IEEEbiography}[{\includegraphics[width=1in,height=1.25in,clip,keepaspectratio]{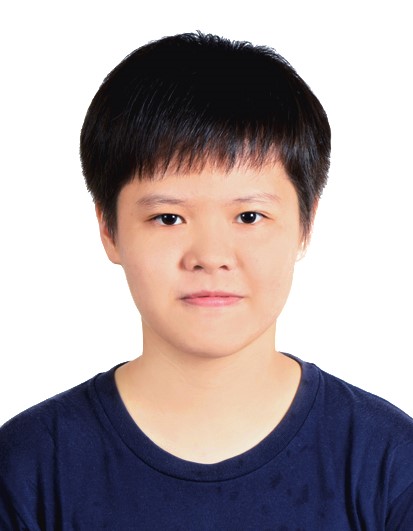}}]{Yi-Hsin Chen} 
received her B.S. degree in applied mathematics from National Chung Hsing University (NCHU), Taiwan, and her M.S. degree in data science and engineering from National Chiao Tung University (NCTU), Taiwan, in 2018 and 2020, respectively. She is currently pursuing her Ph.D. degree in computer science and engineering, National Yang Ming Chiao Tung University (NYCU), Taiwan. Her research interests include learning-based image/video coding, image/video restoration, computer vision, and deep learning.
\end{IEEEbiography}
\vspace{-3em}

\begin{IEEEbiography}[{\includegraphics[width=1in,height=1.25in,clip,keepaspectratio]{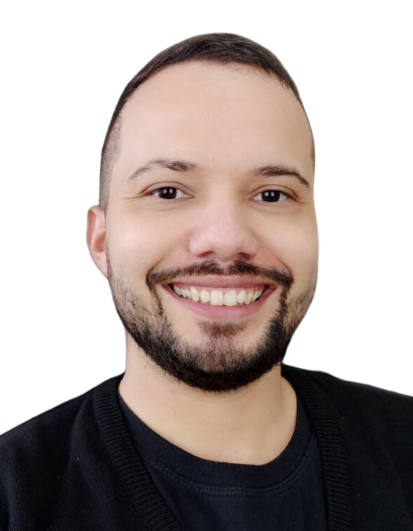}}]{Ruhan Conceição} holds a Master’s degree in Computer Science (2015) and a Bachelor's degree in Computer Engineering (2014) from the Federal University of Pelotas, Brazil, and is currently pursuing a Ph.D. in Computer Science at the same institution. In 2024, he was a visiting Ph.D. researcher at National Yang Ming Chiao Tung University, Taiwan. He is a Professor at the Federal Institute of Education, Science, and Technology of Sul-rio-grandense, where he coordinated the Integrated Technical Program in Internet Informatics from 2019 to 2022. His research interests include hardware design and complexity-aware software solutions for traditional video codecs, light-field imaging, and the quantization of neural network-based video codecs.
\end{IEEEbiography}
\vspace{-3em}

\begin{IEEEbiography}[{\includegraphics[width=1in,height=1.25in,clip,keepaspectratio]{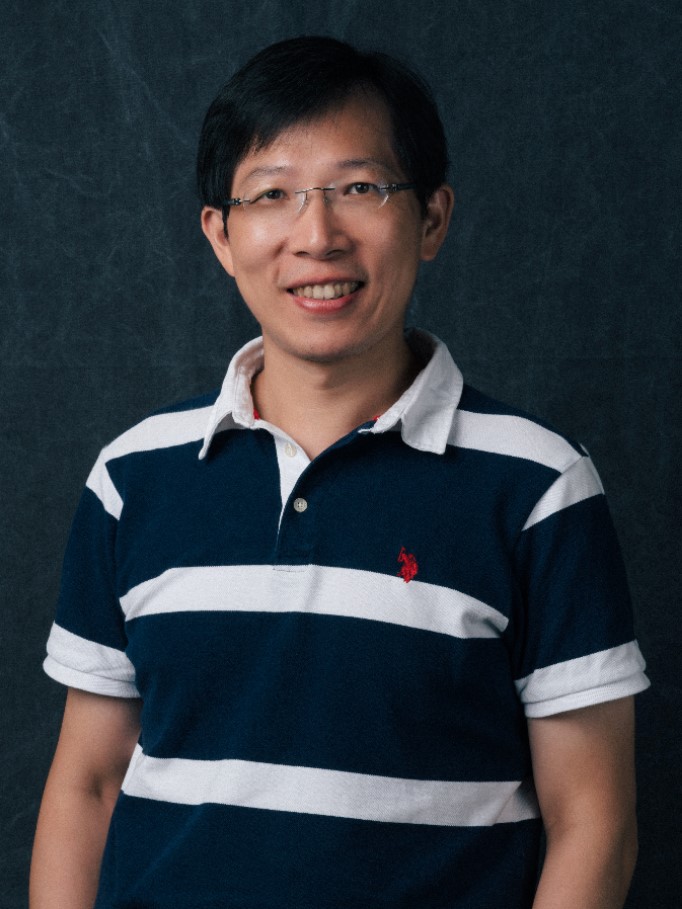}}]{Wen-Hsiao Peng} (M’09-SM’13-F’25) received his Ph.D. degree from National Chiao Tung University (NCTU), Taiwan, in 2005. He was with the Intel Microprocessor Research Laboratory, USA, from 2000 to 2001. Since 2003, he has actively participated in the ISO/IEC and ITU-T video coding standardization process and contributed to the development of H.264/AVC Scalable Amendment, H.265/HEVC, H.265/HEVC Screen Content Coding Extensions (SCC), H.266/VVC, and JPEG AI. He was a Visiting Scholar with the IBM Thomas J. Watson Research Center, USA, from 2015 to 2016. He has authored over 120 journal/conference papers and over 60 ISO/IEC and ITU-T standards contributions. Dr. Peng was Chair of the IEEE Circuits and Systems Society (CASS) Visual Signal Processing (VSPC) Technical Committee from 2020 to 2022. He was Distinguished Lecturer of IEEE CASS (2022-2023) and of APSIPA (2017-2018). He was appointed Editor-in-Chief of the IEEE Journal on Emerging and Selected Topics in Circuits and Systems (JETCAS) for 2024-2025. Dr. Peng is a Fellow of the Higher Education Academy (FHEA), and a Fellow of the IEEE. 
\end{IEEEbiography}
\vspace{-3em}

\begin{IEEEbiography}[{\includegraphics[width=1in,height=1.25in,clip,keepaspectratio]{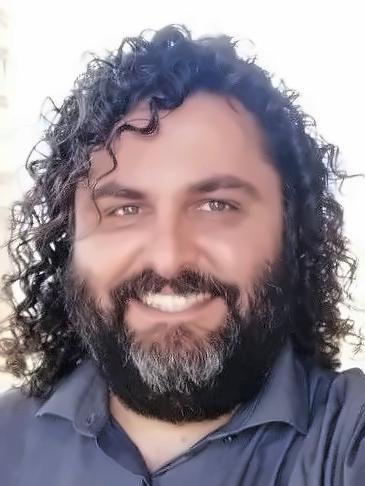}}]{Luciano Volcan Agostini} (Senior Member, IEEE) received his M.S. and Ph.D. degrees in Computer Science from the Federal University of Rio Grande do Sul, Brazil, in 2002 and 2007, respectively. Since 2002, he has been a professor at the Federal University of Pelotas (UFPel), Brazil, where he is currently a Full Professor. Since 2013, he has held the status of CNPq Productivity Research Fellow. He is also the Vice President of the Brazilian Microelectronics Society (SBMicro) and leads the Video Technology Research Group at UFPel. From 2013 to 2017, he served as Vice President for Research and Graduate Studies at UFPel. Additionally, he is a member of the Advisory Committees of CNPq and FAPERGS, two of Brazil’s leading research funding agencies. Dr. Agostini has authored over 300 papers published in prestigious international journals and conferences. He is an Associate Editor for IEEE TCSVT and IEEE OJCS. He is also a Senior Member of ACM and a member of the SBC and SBMicro societies in Brazil.
\end{IEEEbiography}
\vspace{-3em}

\end{document}